\documentclass[letterpaper, 10 pt, conference]{ieeeconf}  

\IEEEoverridecommandlockouts                              
\usepackage{graphics} 
\usepackage{epsfig} 
\usepackage{mathptmx} 
\usepackage{times} 
\usepackage{amsmath} 
\usepackage{amssymb}  
\usepackage{tcolorbox}
\usepackage{algorithm}
\usepackage{placeins}
\usepackage{booktabs}
\usepackage{tabularx}
\usepackage{makecell}
\usepackage{dblfloatfix}
\usepackage{algorithm}
\usepackage{algpseudocode}
\usepackage{url}
\usepackage{hyperref}
\usepackage{tikz}

\newcommand\ieeesubmissiontext{%
    \small\centering
    This work has been submitted to the IEEE for possible publication. \\ Copyright may be transferred without notice, after which this version may no longer be accessible.}

\newcommand\ieeesubmissionnotice{%
  \begin{tikzpicture}[remember picture,overlay]
    \node[anchor=south, yshift=10pt] at (current page.south)
      {\parbox{\dimexpr\textwidth-2\fboxsep\relax}{\ieeesubmissiontext}};
  \end{tikzpicture}}

\title{\LARGE \bf
Karma Mechanisms for Decentralised, Cooperative \\ Multi Agent Path Finding 
}

\author{
    Kevin Riehl$^1$, Julius Schlapbach$^1$, Anastasios Kouvelas, Michail A. Makridis
    \thanks{
    All authors are with Traffic Engineering Group, Institute for Transportation Planning and Systems, ETH Zürich, Stefano Franscini Platz 3, 8053 Zürich, Switzerland.
    $^1$Kevin Riehl and Julius Schlapbach are co-first authors.
    Corresponding author: kriehl@ethz.ch.
    }
}

\begin{document}

\maketitle
\ieeesubmissionnotice
\thispagestyle{empty}
\pagestyle{empty}

\begin{abstract}
Multi-Agent Path Finding (MAPF) is a fundamental coordination problem in large-scale robotic and cyber-physical systems, where multiple agents must compute conflict-free trajectories with limited computational and communication resources. 
While centralised optimal solvers provide guarantees on solution optimality, their exponential computational complexity limits scalability to large-scale systems and real-time applicability. 
Existing decentralised heuristics are faster, but result in suboptimal outcomes and high cost disparities.
This paper proposes a decentralised coordination framework for cooperative MAPF based on Karma mechanisms -- artificial, non-tradeable credits that account for agents’ past cooperative behaviour and regulate future conflict resolution decisions. 
The approach formulates conflict resolution as a bilateral negotiation process that enables agents to resolve conflicts through pairwise replanning while promoting long-term fairness under limited communication and without global priority structures.
The mechanism is evaluated in a lifelong robotic warehouse multi-agent pickup-and-delivery scenario with kinematic orientation constraints. 
The results highlight that the Karma mechanism balances replanning effort across agents, reducing disparity in service times without sacrificing overall efficiency.
Code: 
\href{https://github.com/DerKevinRiehl/karma_dmapf}{https://github.com/DerKevinRiehl/karma\_dmapf}
\end{abstract}

\section{INTRODUCTION}


Multi-robot cyber-physical systems are increasingly deployed in automated environments to perform coordinated tasks, including automated guided vehicles in warehouses, multi-arm assembly systems in manufacturing, coordinated cranes and vehicles at ports, and unmanned aerial vehicle (UAV) swarms~\cite{gautam2012review}.  
In such settings, agents must generate motion plans that avoid collisions with both static obstacles and dynamic entities, such as fellow agents or humans.  
Multi-Agent Path Finding (MAPF) formalises this problem of computing conflict-free trajectories for multiple agents from given start to goal locations on a graph representation of the environment~\cite{gao2024review}.  
Accordingly, MAPF has attracted significant research interest across artificial intelligence, operations research, robotics, and control theory.

Depending on the application, previous works on MAPF differ in their discretisation of space and time, objective functions (e.g. minimizing makespan vs. sum-of-costs), dimensionality (2D vs. 3D), constraints (kinematics, obstacles, task priorities), and simplifying assumptions regarding agent heterogeneity, execution delay, and task assignment~\cite{stern2019multi,yamauchi2021path}.

Several optimal solvers for the MAPF problem have been proposed that are guaranteed to find an optimal solution, if it exists. These can be grouped into four families: (i)~reduction-based, (ii)~A*-based, (iii)~ICTS-based, and (iv)~CBS-based approaches. 
\textit{Reduction-based} solvers formulate MAPF as well-studied operations research problems, such as flow networks and integer linear programming~\cite{yu2016optimal}, or branch-and-cut-and-price formulations~\cite{lam2022branch}. These differ fundamentally from the other three search-based approaches.
\textit{A*-based} methods~\cite{hart1968formal} explicitly construct and explore the joint state space of all agents using heuristic search, where each node represents a global configuration of all agents and the heuristic guides expansion towards states with lower estimated costs.
\textit{ICTS-based} (Increasing Cost Tree Search) methods~\cite{sharon2013increasing} explore a tree whose nodes represent cost vectors for the individual agents' trajectories. Since not every cost vector admits a conflict-free joint solution, the algorithm iteratively expands the tree by increasing individual costs until a feasible combination of trajectories is identified.
\textit{CBS-based} (Conflict-Based Search)~\cite{sharon2012meta} methods perform a two-level search: a high-level search over a binary constraint tree that branches on detected conflicts by adding pairwise constraints to agents, combined with low-level replanning of individual agents under accumulating constraints.
Both ICTS and CBS typically invoke A* (or other shortest-path algorithms) to find optimal single-agent trajectories under accumulated constraints, but avoid constructing and exploring the complete time-expanded state space.
While these approaches find globally optimal solutions, they rely on centralised computation and full observability, limiting their applicability in real-time and large-scale systems.  
Furthermore, the combinatorial growth of the joint state space renders these methods computationally challenging, highlighting the need for scalable control architectures.

These limitations have motivated the study of decentralised MAPF (DMAPF), where agents make decisions based on a subset of the state space only~\cite{wang2020walk}.  
Decentralised approaches align naturally with distributed control paradigms, where global coordination emerges from local interaction rules.  
Existing work considers both cooperative~\cite{purwin2008theory} and competitive settings~\cite{amir2015multi}.  
Additionally, approaches distinguish between offline planning and online conflict resolution, with the latter being essential for adaptive control in dynamic environments~\cite{maoudj2022decentralized,farhadi2025evolution}.  
Some methods rely on explicit communication protocols~\cite{farhadi2025evolution}, while others employ reservation-based schemes~\cite{inotsume2020path} or implicit coordination through behavioural prediction~\cite{mavrogiannis2020decentralized}.  
In a growing branch of research, such decentralised coordination is discoursed using economic terminology, such as negotiations~\cite{ho2020decentralized,keskin2024decentralized}, sequential bargaining processes~\cite{pritchett2017negotiated}, and more formally as combinatorial auctions~\cite{amir2015multi}.
Auction-based negotiations for MAPF can make use of artificial currencies such as merit-based tokens~\cite{keskin2024decentralized,desaraju2012decentralized}, which motivates dedicated (incentive) mechanism designs~\cite{chandra2023socialmapf}.
Especially in non-cooperative settings, economic mechanisms can enable incentive compatibility through privacy~\cite{keskin2024decentralized}, and support prioritisation and fairness considerations~\cite{ho2020decentralized,chandra2023socialmapf}.
Unlike optimal centralised solvers, decentralised heuristics provide no optimality and robustness guarantees, due to limited local information.
Karma mechanisms represent a class of artificial currency schemes based on non-tradeable credits~\cite{riehl2024resource}.  
Originally designed for resource allocation in peer-to-peer networks~\cite{vishnumurthy2003karma}, they have been applied to coordination problems in a variety of domains, including cloud computing, telecommunication, road transportation, and social assignment and negotiation problems.
A key property of Karma mechanisms is their ability to incentivise cooperative behaviour in decentralised systems, even among self-interested agents.  
By linking present actions to future earning potential, they promote forward-looking decision-making among agents.
From a control perspective, they can be interpreted as distributed feedback mechanisms that shape agent decisions through economic incentives rather than explicit commands.  
This makes them particularly suitable for large-scale systems where direct coordination is infeasible or undesirable.

In this work, we propose a Karma mechanism-based approach to DMAPF that embeds incentive dynamics directly into decentralised conflict resolution.
Our framework formulates trajectory coordination as a distributed control problem with endogenous priority adaptation driven by agents' cooperation history.
In particular, Karma acts as an integral feedback signal that promotes fairness while preserving scalability and responsiveness in online coordination, and promotes long-term indirect reciprocity amongst agents~\cite{riehl2024resource}. 

We demonstrate the effectiveness of the proposed mechanism through an orientation-aware, lifelong, multi-agent pickup-and-delivery case study in a robotic warehouse-like environment. 
The results show that Karma-based coordination achieves efficiency comparable to established decentralised heuristics such as token-passing and negotiation-based approaches with egoistic or altruistic policies, while significantly reducing disparities in service time across tasks. 
These findings establish that incentive-based feedback mechanisms provide a principled approach to fair and scalable coordination in decentralised multi-agent control systems.

The source code and related materials are available on GitHub: \href{https://github.com/DerKevinRiehl/karma_dmapf}{https://github.com/DerKevinRiehl/karma\_dmapf}.


\section{PROBLEM FORMULATION \& REVIEW}
\begin{figure} [!ht]
    \centering
    \includegraphics[width=\linewidth]{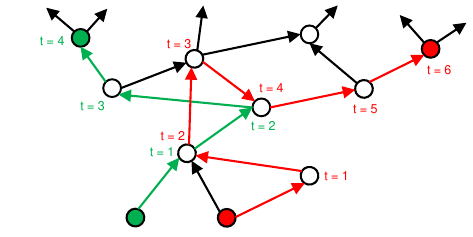}
    \caption{\textbf{Multi Agent Path Finding Problem (MAPF).}}
    \label{fig:mapf_problem}
\end{figure}

MAPF is the problem of computing conflict-free trajectories for agents from start to goal locations~\cite{keskin2024decentralized} (see Fig.~\ref{fig:mapf_problem}).

\subsection{Definitions \& Notation}
Consider $n$ agents navigating through an environment represented as a graph $G=(V,E)$, with start locations $s_i \in V$ and goals $g_i \in V$.
A trajectory $\pi_i = [(s_i,t_0), \dots , (g_i, t_N)]$ for agent $i$ is a sequence of states, where each state is a tuple consisting of vertex $v \in V$ and discrete time step $t$.
The cost of $\pi_i$ can be chosen as an arbitrary application-specific function $c: \Pi \rightarrow \mathbb{R}_{\geq 0}$, where $\Pi$ denotes the set of all possible trajectories.
In this work, we use $c(\pi_i) = N$, the trajectory length in time steps.
Two trajectories $\pi_i$ and $\pi_j$ are considered conflict-free if, for any time step $t \in \{0,\dots,T \}$, the two agents neither occupy the same vertex simultaneously (vertex conflict) nor traverse the same edge at the same time (edge conflict).
An agent is considered \textit{idle} after reaching its goal and before being assigned a new destination.

\subsection{Centralised MAPF with CBS}

CBS is a widely-used, optimal, centralised algorithm for solving MAPF~\cite{sharon2012meta}.  
It decomposes the problem into global coordination for conflict resolution and local trajectory generation.  
CBS maintains a high-level search tree, referred to as the constraint tree $CT$, where each leaf encodes a set of constraints and a corresponding valid solution.  
A constraint is defined as a tuple $(i, v, t)$ or $(i, e, t)$, prohibiting agent~$i$ from occupying vertex~$v$ or traversing edge~$e$ at time~$t$.  
At each node of the $CT$, a low-level planner computes an individually optimal trajectory $\pi_i$ for each agent subject to the accumulated constraints.  
This low-level trajectory planning step is typically performed using A* search on the time-expanded graph.  
The resulting set of trajectories is then checked for conflicts.
If no conflicts are detected at the lowest-cost leaf node, the current solution is globally feasible and optimal, and CBS terminates.  
Otherwise, a conflict between two agents is resolved by branching the $CT$, generating two child nodes with an added constraint for either of the conflicting agents.  
The high-level search proceeds by exploring the $CT$, using best-first search ordered by global cost.  
CBS is complete and cost-optimal but has an exponential worst-case time complexity in the number of agents and conflicts, motivating decentralised alternatives.  

\subsection{Decentralised MAPF Heuristics}

\subsubsection{Token-Passing}

Token-passing approaches establish a sequential planning order among agents, where a shared token grants planning priority.  
Only the agent holding the token updates its trajectory, while all other agents treat their plans as fixed~\cite{stern2019multi}. 

Let $\mathcal{A}_{\text{planned}} \subset \{1,\dots,n\}$ denote the set of agents with committed trajectories.  
An agent $i \notin \mathcal{A}_{\text{planned}}$ computes its trajectory $\pi_i$ by solving a shortest-path problem on a time-expanded graph that avoids all reserved vertices and edges induced by $\{\pi_j\}_{j \in \mathcal{A}_{\text{planned}}}$.  

While this mechanism simplifies coordination, it is limited to sequential decision-making and leads to inefficiencies due to the neglect of interrelations in the joint state space. 
Moreover, the planning order creates asymmetry, with later-planning agents facing more constraints, resulting in globally suboptimal solutions.

\begin{algorithm}
\caption{Negotiation-Based Coordination for MAPF}\label{alg:negotiation_algo}
\begin{algorithmic}[1]
\State $\mathcal{A}_{\text{considered}}^i$ = [] \Comment{Initialize empty list}
\State $\pi_i$ = A*$[ \mathcal{A}_{\text{considered}}^i ]$ \Comment{Plan candidate trajectory}
\State $\pi_i \rightarrow \mathcal{C}_i$ \Comment{Determine conflicts}
\While{$| \mathcal{C}_i | > 0$} \vspace{0.3em}
    \State Select highest priority conflict $c_{i,j,t}^p$
        \State \hspace{\algorithmicindent} $c_{i,j,t}^p = \arg\max_{c_{i,j,t} \in \mathcal{C}_i} \Delta_i^{(j)}$ \vspace{0.3em}
    \State Negotiate conflict resolution responsibility
        \State \hspace{\algorithmicindent} $\pi_i, \pi_j \rightarrow \Delta_i, \Delta_j$ \Comment{Determine costs}
        \State \hspace{\algorithmicindent} $r = \mathcal{N}(\Delta_i,\Delta_j) \in \{i,j\}$ 
        \State \hspace{\algorithmicindent} $\mathcal{A}^r_{\text{considered}} \leftarrow \mathcal{A}^r_{\text{considered}} \cup \{ \bar{r} \}$  
        \State \hspace{\algorithmicindent} $\pi_i$ = A*$[ \mathcal{A}_{\text{considered}}^i ]$\Comment{Replanning}
    \State $\pi_i \rightarrow \mathcal{C}_i$ \Comment{Recompute conflicts}
\EndWhile
\end{algorithmic}
\end{algorithm}

\subsubsection{Negotiation-Based Coordination}
To overcome the limitations of static, fixed-order planning, agents can engage in local negotiations for pairwise conflict resolution through dynamic replanning.
In this setting, each agent initially computes its individually optimal trajectory, ignoring others, and subsequently resolves conflicts iteratively through pairwise interaction, until all trajectories are conflict-free~\cite{gao2024review}  (Alg.~\ref{alg:negotiation_algo}). If no solution is found for an agent, it remains at its current position until the next time step.

Let $\pi_i$ denote the current candidate trajectory of agent $i$, computed while avoiding the time-expanded trajectories of agents in the subset $\mathcal{A}_{\text{considered}}^i$ (initially empty).
The set of conflicts between $\pi_i$ and other agents’ trajectories is denoted as $\mathcal{C}_i$, where each conflict $c_{i,j,t}\in\mathcal{C}_i$ represents a vertex or edge conflict with agent $j$ at a specific time step~$t$.
If $\pi_i$ is conflict-free ($| \mathcal{C}_i | = 0$), the planning for $i$ is terminated.  
Otherwise, agent $i$ selects a conflict $c^p_{i,j,t} \in \mathcal{C}_i$ according to a priority rule based on deviation cost: 
\begin{equation}
    c^p_{i,j,t} = \arg\max_{c_{i,j,t} \in \mathcal{C}_i} \Delta_i^{(j)}
\end{equation}
where $\Delta_i^{(j)}$  is the estimated trajectory cost increase for agent~$i$ when replanning to avoid conflicting agent~$j$.

Agent $i$ then initiates a bilateral negotiation with agent $j$ to resolve the conflict.
Specifically, each agent computes an alternative trajectory that avoids the conflict and determines the associated cost increase relative to their current candidate trajectories, denoted by the shorthand notation $\Delta_i = \Delta_i^{(j)}$ and $\Delta_j = \Delta_j^{(i)}$ for agents $i$ and $j$, respectively.  
A negotiation mechanism $\mathcal{N}(\Delta_i,\Delta_j)\rightarrow r \in \{i,j\}$ then determines which agent $r$ has to adapt their trajectory.

In the egoistic setting, agent $j$ agrees to replan only if it incurs non-negative incremental cost, reflecting a purely self-interested decision rule.  
Formally,  
\begin{equation}
\label{eq:egoistic_negotiation_rule}
    \mathcal{N}_{\text{ego}}(\Delta_i,\Delta_j) =
    \begin{cases}
    j, & \text{if } \Delta_j \leq 0, \\
    i, & \text{otherwise}.
    \end{cases}
\end{equation}

In the altruistic setting, agent $j$ agrees to change its trajectory if its cost increase is smaller than agent $i$'s, reflecting cooperative behaviour that seeks to minimise overall costs.  
Formally,  
\begin{equation}
\label{eq:altruistic_negotiation_rule}
    \mathcal{N}_{\text{alt}}(\Delta_i,\Delta_j) =
    \begin{cases}
    j, & \text{if } \Delta_j < \Delta_i, \\
    i, & \text{if } \Delta_j > \Delta_i, \\
    \mathcal{U}(\{i, j\}), & \text{if } \Delta_i = \Delta_j
    \end{cases}
\end{equation}

\noindent In the case $\Delta_i = \Delta_j$, the assignment is resolved by random tie-breaking with uniform probabilities, denoted by $\mathcal{U}(\cdot)$.  

After the negotiation, the selected agent $r \in \{ i, j \}$ replans its trajectory, while avoiding the other agent $\bar{r}= \{ i,j \} \setminus r$.
\begin{equation}
    \mathcal{A}^r_{\text{considered}} \leftarrow \mathcal{A}^r_{\text{considered}} \cup \{ \bar{r} \}.
\end{equation}
\noindent Agent $i$ iterates this process of conflict detection and negotiation, until $\pi_i$ is conflict-free ($| \mathcal{C}_i | = 0$).

\begin{figure} [!t]
    \centering
    \includegraphics[width=0.9\linewidth]{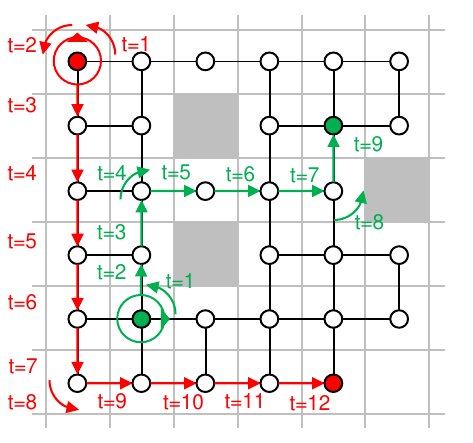}
    \caption{\textbf{Warehouse MAPD Case Study with Kinematic Constraints.}}
    \label{fig:case_study}
\end{figure}

\section{METHODS}
\subsection{Karma-based Negotiation Mechanism}

To regulate decentralised interactions beyond local cost comparisons, we introduce a Karma-based negotiation mechanism.
Each agent $i$ maintains a state variable $k_i \in \mathbb{Z}$ referred to as its \textit{Karma} balance, which evolves over time based on past decisions.
Depending on the application, the balance may be subject to redistribution mechanisms across agents or regular event-based resets.
Karma can be interpreted as an internal credit that encodes an agent’s history of cooperation.  
From a control perspective, $k_i$ acts as an auxiliary state variable that introduces temporal coupling between decisions, thereby shaping future behaviour through feedback.  

\begin{figure*}[!t]
    \centering
    \includegraphics[width=\linewidth]{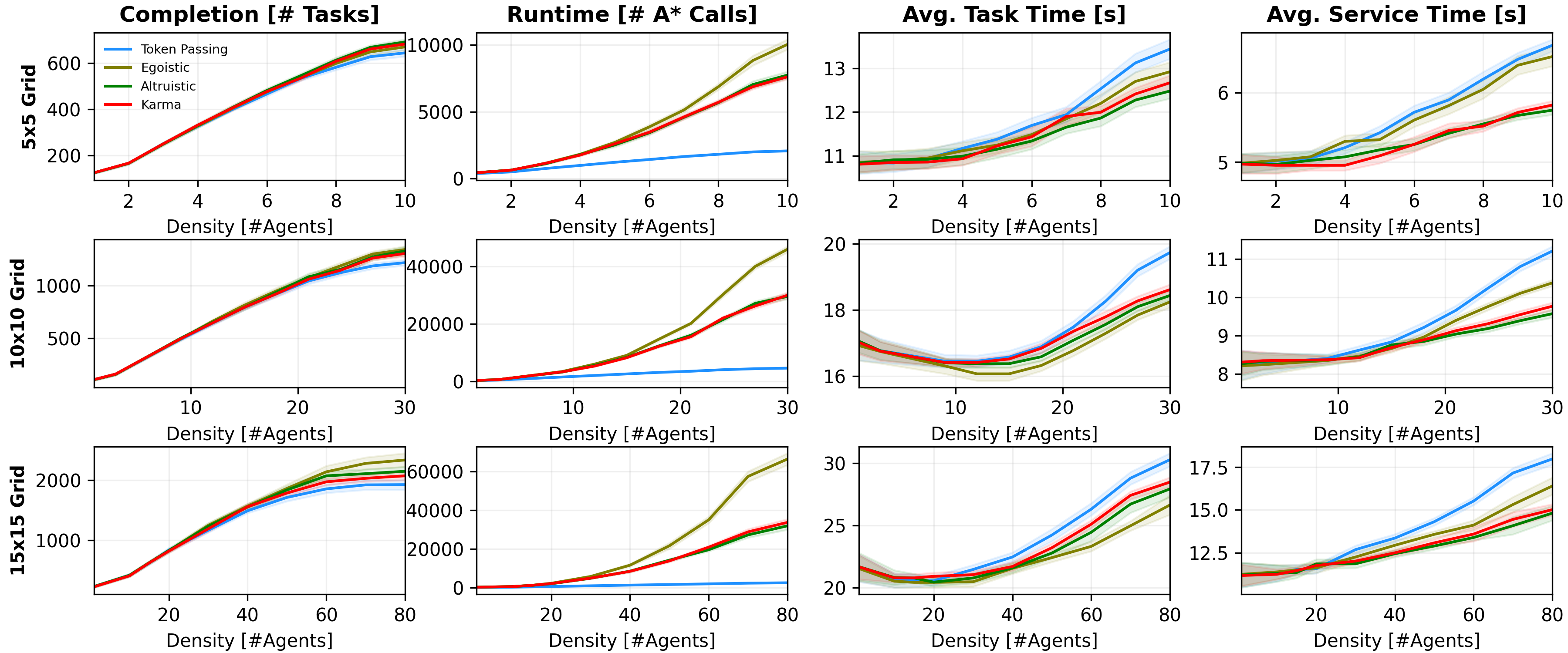}
    \caption{\textbf{Warehouse MAPD Case Study Benchmark.}}
    \label{fig:efficiency_benchmark}
\end{figure*} 

Given a conflict between agents $i$ and $j$, the Karma-based negotiation mechanism is formally defined as:
\begin{equation}
\label{eq:karma_negotiation_mechanism}
    \mathcal{N}_{\text{karma}}(\Delta_i,\Delta_j, k_i, k_j) =
    \begin{cases}
    j, & \text{if } \Delta_j + \tau k_j < \Delta_i + \tau k_i, \\
    i, & \text{if } \Delta_j + \tau k_j > \Delta_i + \tau k_i, \\
    \mathcal{U}(\{i, j\}), & \text{otherwise},
    \end{cases}
\end{equation}
where $\tau \geq 0$ is a design parameter weighing the influence of the agent's current Karma balance.

This formulation biases the replanning decision towards agents with lower Karma, effectively rewarding past cooperative behaviour.  
After negotiation, the Karma balances are updated as follows, where $r = \mathcal{N}_{\text{karma}}(\Delta_i,\Delta_j, k_i, k_j)$ is the replanning agent.
\begin{equation}
k_r \leftarrow k_r + \Delta_{r}, \quad
k_{\bar{r}} \leftarrow k_{\bar{r}} - \Delta_{r}.
\end{equation}

This update rule ensures that agents that replan in conflicts accumulate Karma, while agents whose paths remain unchanged compensate them.  
As a result, agents that have replanned frequently in the past are associated with an increased composite cost in subsequent negotiations, thereby balancing the replanning burden among agents over time.
The mechanism can thus be interpreted as a distributed integral feedback controller that regulates long-term indirect reciprocity in decentralised decision-making.

\subsection{Simulation Case Study}
We demonstrate the performance of the proposed, Karma-based DMAPF using the example of a lifelong, orientation-aware, stay-at-target, synchronous, multi-agent pickup and delivery (MAPD) setup in a discrete-time and discrete-space, robotic warehouse-like environment, as shown in Fig.~\ref{fig:case_study}.
During the simulation, randomly generated tasks are assigned to the closest available (or soon to be available) agent, using linear sum assignment~\cite{kuhn1955hungarian}.
Agents navigate to their assigned task's start location, pick up the task, navigate to its goal, and deliver the task.
The environment's graph $G$ represents a square grid, where agents can move to their Von Neumann neighbourhood, while complying with their orientation-aware kinematic constraints. 
Each simulation runs for 100 time steps and is repeated for multiple random seeds, to calculate mean and standard deviation of the evaluation metrics, which include the number of completed tasks, the average cost per task, the dispersion of task costs (standard deviation), and runtime (number of A* calls).
The Karma-based DMAPF is evaluated on three different grid sizes of $5\times5$, $10\times10$, and $15\times15$ cells and varying agent counts. Tasks are randomly spawned within the grid, which is surrounded by a one-cell-wide border that agents can also traverse. Trajectories are planned on a space-time graph with kinematic (orientation-aware) constraints, where the cost of a path refers to the number of time steps it takes for execution.
The Karma balance of each agent is reset to $k_i = 0$ upon pickup of a new task.


\section{RESULTS}

\begin{figure*} [!t]
    \centering
    \includegraphics[width=\linewidth]{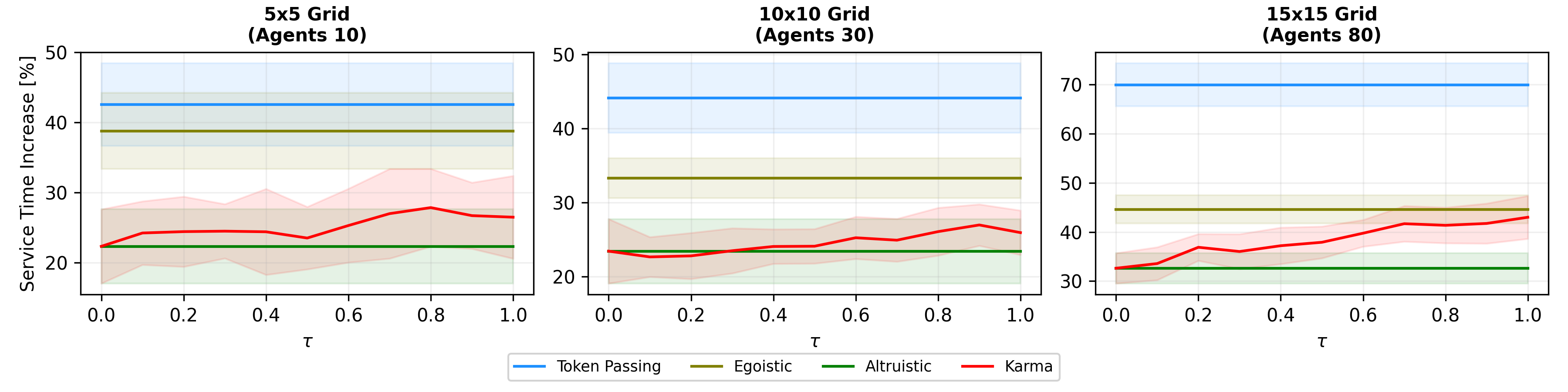}
    \caption{\textbf{Impact of the Karma Influence Parameter on the Increase in Task Service Time.}}
    \label{fig:karma_parameter_influence}
\end{figure*}

\subsection{General Overview}

We compare the proposed Karma-based negotiation mechanism against the token-passing approach, as well as egoistic and altruistic negotiation policies (Eqs.~\eqref{eq:egoistic_negotiation_rule}, \eqref{eq:altruistic_negotiation_rule}). Fig.~\ref{fig:efficiency_benchmark} illustrates performance in lifelong MAPD scenarios across three grid sizes ($5 \times 5$, $10 \times 10$, $15 \times 15$). We measure the \textit{task time} between the assignment of a task to an agent and its delivery at the goal, and the \textit{service time} between pickup and delivery only. \par

Token-passing consistently completes the fewest tasks due to its sequential planning structure and the inability to adapt previously committed trajectories. This forces later-planning agents to avoid any conflicts, resulting in longer routes and higher task completion times. Although the number of A* calls is directly proportional to the number of assigned tasks, this computational efficiency is achieved by sacrificing solution efficiency, as evidenced by the increased task and service times. \par

Introducing the possibility for agents to negotiate the pairwise conflict resolution responsibility improves overall average performance with more completed tasks and reduced average costs, at the expense of additional A* calls to compute alternative trajectories and associated deviation costs. Despite less frequent trajectory changes by agents with committed solutions under the egoistic negotiation rule, this approach requires more A* calls than the other considered negotiation-based approaches. Since $\Delta_i$ and $\Delta_j$ implicitly include future conflict avoidance efforts with respect to agents in $\mathcal{A}^i_{\text{considered}}$ (for $i$) or all other agents with committed trajectories (for $j$), the altruistic and Karma-based policies guide the system toward solutions requiring fewer agent negotiations and associated graph searches. \par

The altruistic negotiation rule compares the anticipated cost increases $\Delta_i$ and $\Delta_j$ for the respective agents under the responsibility for the conflict resolution. Since the cost equals the trajectory length from the current position to the goal, agents with longer remaining distances tend to win negotiations against agents with shorter trajectories. In contrast, the egoistic policy makes no such distinctions. Correspondingly, since the distance between an agent's current position and the pickup location of an assigned task is typically smaller than the average service trajectory length from pickup to delivery in our simulation scenario (due to the linear sum assignment), the average performance advantage of the altruistic over the egoistic policy is reduced when considering the entire task duration.


\subsection{Effect of Karma Influence Parameter $\tau$}

The Karma-based negotiation mechanism (Eq. \eqref{eq:karma_negotiation_mechanism}) weighs immediate replanning costs against accumulated Karma for each agent with the influence parameter $\tau \in \mathbb{R}_{\geq 0}$. At $\tau = 0$, the mechanism reduces to the altruistic policy. Increasing $\tau$ shifts priority from immediate replanning cost increases towards balancing accumulated long-term costs across agents. \par

Fig. \ref{fig:karma_parameter_influence} illustrates the effect of $\tau$ on the mean task service time relative to the shortest possible path without considering other agents, for a simulation with 100 time steps (shaded region indicates the standard deviation over $20$ random seeds). As the influence of the Karma balance increases, agents prioritize the equal distribution of conflict resolution efforts over immediate cost minimization, demonstrating the trade-off between the average incurred cost and its spread across tasks. \par

At low $\tau$ values, the mechanism behaves similarly to the altruistic policy with inconsistent impact, while Karma dominates the negotiation for large $\tau$ values over the immediate cost, resulting in higher service times. Based on these empirical results, $\tau = 0.5$ proved as a suitable choice for subsequent experiments, as it achieves a balance between cost efficiency and fairness. 
\par

\subsection{Efficiency and Variance Implications of Karma}

While the egoistic and altruistic negotiation rules consider immediate deviation costs, they do not explicitly regulate how replanning effort is distributed across agents over time. In contrast, the Karma-based negotiation mechanism introduces an additional feedback signal that accumulates past cooperation behaviour and influences future conflict resolution decisions. As a result, Karma enables a more balanced allocation of coordination effort across tasks.

\begin{figure} [!t]
    \centering
    \includegraphics[width=\linewidth]{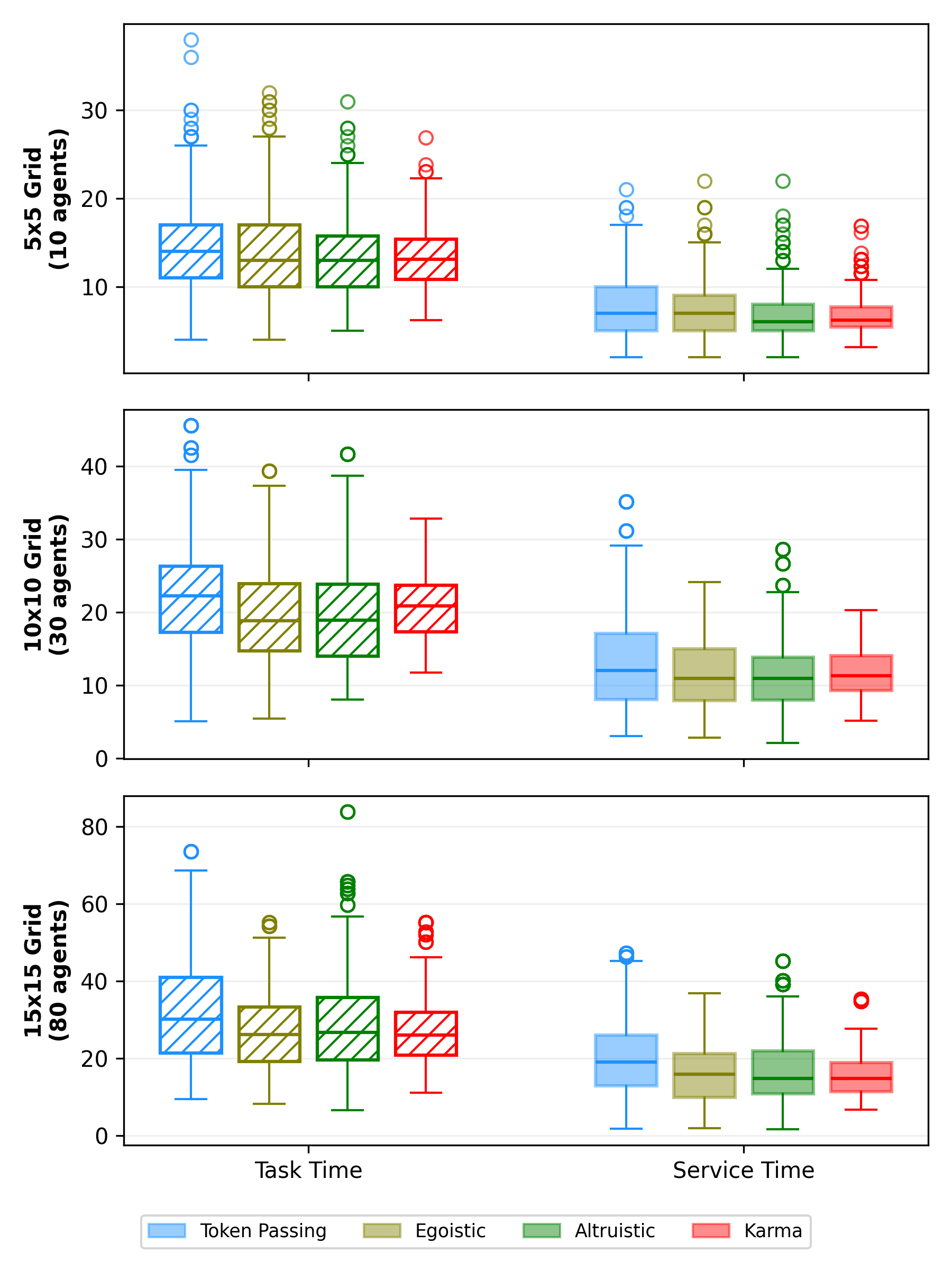}
    \caption{\textbf{Task and Service Time Distribution.}}
    \label{fig:delay_distribution}
\end{figure}

Fig.~\ref{fig:delay_distribution} compares the dispersion of absolute task and service times across negotiation strategies, reporting the median, interquartile range and outliers in a box plot diagram. Although the Karma-based mechanism achieves efficiency levels comparable to egoistic and altruistic negotiation in terms of average task and service time, it consistently reduces their dispersion. This indicates that Karma promotes a more equitable distribution of delays without sacrificing overall system throughput.




\section{CONCLUSIONS}

In this study, we proposed a Karma-based decentralised coordination mechanism for MAPF, in which bilateral conflict resolution is augmented with an artificial credit balance encoding agents' past cooperative behaviour. 
The resulting framework interprets decentralised trajectory coordination as a distributed control problem with endogenous priority adaptation, where Karma acts as an integral feedback signal to balance long-term replanning burden. 
In a lifelong, orientation-aware MAPD case study, the proposed mechanism achieved efficiency comparable to established decentralised heuristics, including token-passing and negotiation-based approaches with egoistic and altruistic rules, while reducing disparities in task and service times across tasks. 
These results demonstrate that incentive-based feedback can improve fairness in decentralised multi-agent coordination without sacrificing scalability or requiring centralised optimisation.
More broadly, the presented results suggest that artificial-currency feedback mechanisms are a promising direction for fair, scalable, and adaptive coordination in decentralised multi-agent control systems.

The Karma update rule and its influence parameter $\tau$ were studied empirically, and their behaviour may depend on the problem setup, task density, and interaction structure. 
Moreover, the current study focuses primarily on task and service time statistics, whereas other notions of fairness, robustness, and communication overhead deserve further analysis.
Future work therefore might address an analysis of stability, convergence, and performance bounds of Karma-based negotiation. 
Furthermore, the impact of different Karma payment rules (pay-to-peer vs. pay-to-society), limitation of Karma balances (minimum and maximum Karma), and redistribution or resetting schemes remain to be explored.

\bibliographystyle{IEEEtran}
\bibliography{references}

@inproceedings{gautam2012review,
  title={A review of research in multi-robot systems},
  author={Gautam, Avinash and Mohan, Sudeept},
  booktitle={2012 IEEE 7th international conference on industrial and information systems (ICIIS)},
  pages={1--5},
  year={2012},
  organization={IEEE},
  doi={10.1109/ICIInfS.2012.6304778}
}

@article{gao2024review,
  title={A review of graph-based multi-agent pathfinding solvers: From classical to beyond classical},
  author={Gao, Jianqi and Li, Yanjie and Li, Xinyi and Yan, Kejian and Lin, Ke and Wu, Xinyu},
  journal={Knowledge-Based Systems},
  volume={283},
  pages={111121},
  year={2024},
  publisher={Elsevier},
  doi={10.1016/j.knosys.2023.111121}
}

@article{stern2019multi,
  title={Multi-agent path finding--an overview},
  author={Stern, Roni},
  journal={Artificial intelligence: 5th RAAI summer School, dolgoprudny, Russia, july 4--7, 2019, tutorial lectures},
  pages={96--115},
  year={2019},
  publisher={Springer}
}

@inproceedings{sharon2012meta,
  title={Meta-agent conflict-based search for optimal multi-agent path finding},
  author={Sharon, Guni and Stern, Roni and Felner, Ariel and Sturtevant, Nathan},
  booktitle={Proceedings of the International Symposium on Combinatorial Search},
  volume={3},
  number={1},
  pages={97--104},
  year={2012}
}

@article{sharon2013increasing,
  title={The increasing cost tree search for optimal multi-agent pathfinding},
  author={Sharon, Guni and Stern, Roni and Goldenberg, Meir and Felner, Ariel},
  journal={Artificial intelligence},
  volume={195},
  pages={470--495},
  year={2013},
  publisher={Elsevier}
}

@article{yu2016optimal,
  title={Optimal multirobot path planning on graphs: Complete algorithms and effective heuristics},
  author={Yu, Jingjin and LaValle, Steven M},
  journal={IEEE Transactions on Robotics},
  volume={32},
  number={5},
  pages={1163--1177},
  year={2016},
  publisher={IEEE}
}

@article{lam2022branch,
  title={Branch-and-cut-and-price for multi-agent path finding},
  author={Lam, Edward and Le Bodic, Pierre and Harabor, Daniel and Stuckey, Peter J},
  journal={Computers \& Operations Research},
  volume={144},
  pages={105809},
  year={2022},
  publisher={Elsevier}
}

@article{hart1968formal,
  title={A formal basis for the heuristic determination of minimum cost paths},
  author={Hart, Peter E and Nilsson, Nils J and Raphael, Bertram},
  journal={IEEE transactions on Systems Science and Cybernetics},
  volume={4},
  number={2},
  pages={100--107},
  year={1968},
  publisher={IEEE}
}

@article{purwin2008theory,
  title={Theory and implementation of path planning by negotiation for decentralized agents},
  author={Purwin, Oliver and D’Andrea, Raffaello and Lee, Jin-Woo},
  journal={Robotics and Autonomous Systems},
  volume={56},
  number={5},
  pages={422--436},
  year={2008},
  publisher={Elsevier}
}

@article{desaraju2012decentralized,
  title={Decentralized path planning for multi-agent teams with complex constraints},
  author={Desaraju, Vishnu R and How, Jonathan P},
  journal={Autonomous Robots},
  volume={32},
  number={4},
  pages={385--403},
  year={2012},
  publisher={Springer}
}

@inproceedings{amir2015multi,
  title={Multi-agent pathfinding as a combinatorial auction},
  author={Amir, Ofra and Sharon, Guni and Stern, Roni},
  booktitle={Proceedings of the AAAI Conference on Artificial Intelligence},
  volume={29},
  number={1},
  year={2015}
}

@article{pritchett2017negotiated,
  title={Negotiated decentralized aircraft conflict resolution},
  author={Pritchett, Amy R and Genton, Antoine},
  journal={IEEE transactions on intelligent transportation systems},
  volume={19},
  number={1},
  pages={81--91},
  year={2017},
  publisher={IEEE}
}

@inproceedings{inotsume2020path,
  title={Path negotiation for self-interested multirobot vehicles in shared space},
  author={Inotsume, Hiroaki and Aggarwal, Aayush and Higa, Ryota and Nakadai, Shinji},
  booktitle={2020 IEEE/RSJ International Conference on Intelligent Robots and Systems (IROS)},
  pages={11587--11594},
  year={2020},
  organization={IEEE}
}

@inproceedings{maoudj2022decentralized,
  title={Decentralized multi-agent path finding in warehouse environments for fleets of mobile robots with limited communication range},
  author={Maoudj, Abderraouf and Christensen, Anders Lyhne},
  booktitle={International Conference on Swarm Intelligence},
  pages={104--116},
  year={2022},
  organization={Springer}
}

@inproceedings{mavrogiannis2020decentralized,
  title={Decentralized multi-agent navigation planning with braids},
  author={Mavrogiannis, Christoforos I and Knepper, Ross A},
  booktitle={Algorithmic Foundations of Robotics XII: Proceedings of the Twelfth Workshop on the Algorithmic Foundations of Robotics},
  pages={880--895},
  year={2020},
  organization={Springer}
}

@article{wang2020walk,
  title={Walk, stop, count, and swap: decentralized multi-agent path finding with theoretical guarantees},
  author={Wang, Hanlin and Rubenstein, Michael},
  journal={IEEE Robotics and Automation Letters},
  volume={5},
  number={2},
  pages={1119--1126},
  year={2020},
  publisher={IEEE}
}

@inproceedings{yamauchi2021path,
  title={Path and action planning in non-uniform environments for multi-agent pickup and delivery tasks},
  author={Yamauchi, Tomoki and Miyashita, Yuki and Sugawara, Toshiharu},
  booktitle={European Conference on Multi-Agent Systems},
  pages={37--54},
  year={2021},
  organization={Springer}
}

@article{ho2020decentralized,
  title={Decentralized multi-agent path finding for UAV traffic management},
  author={Ho, Florence and Geraldes, R{\'u}ben and Gon{\c{c}}alves, Artur and Rigault, Bastien and Sportich, Benjamin and Kubo, Daisuke and Cavazza, Marc and Prendinger, Helmut},
  journal={IEEE Transactions on Intelligent Transportation Systems},
  volume={23},
  number={2},
  pages={997--1008},
  year={2020},
  publisher={IEEE}
}

@article{chandra2023socialmapf,
  title={Socialmapf: Optimal and efficient multi-agent path finding with strategic agents for social navigation},
  author={Chandra, Rohan and Maligi, Rahul and Anantula, Arya and Biswas, Joydeep},
  journal={IEEE Robotics and Automation Letters},
  volume={8},
  number={6},
  pages={3214--3221},
  year={2023},
  publisher={IEEE}
}

@article{keskin2024decentralized,
  title={Decentralized multi-agent path finding framework and strategies based on automated negotiation},
  author={Keskin, M Onur and Cant{\"u}rk, Furkan and Eran, Cihan and Aydo{\u{g}}an, Reyhan},
  journal={Autonomous Agents and Multi-Agent Systems},
  volume={38},
  number={1},
  pages={10},
  year={2024},
  publisher={Springer}
}

@article{farhadi2025evolution,
  title={Evolution of path costs for efficient decentralized multi-agent pathfinding},
  author={Farhadi, Ulrich and Hess, Henning and Maoudj, Abderraouf and Christensen, Anders Lyhne},
  journal={Swarm and Evolutionary Computation},
  volume={93},
  pages={101833},
  year={2025},
  publisher={Elsevier}
}

@article{riehl2024resource,
  title={Resource Allocation with Karma Mechanisms—A Review},
  author={Riehl, Kevin and Kouvelas, Anastasios and Makridis, Michail A},
  journal={Economies},
  volume={12},
  number={8},
  pages={211},
  year={2024},
  publisher={MDPI}
}

@inproceedings{vishnumurthy2003karma,
  title={Karma: A secure economic framework for peer-to-peer resource sharing},
  author={Vishnumurthy, Vivek and Chandrakumar, Sangeeth and Sirer, Emin Gun},
  booktitle={Workshop on Economics of Peer-to-peer Systems},
  volume={35},
  number={6},
  year={2003}
}

@article{kuhn1955hungarian,
  title={The Hungarian method for the assignment problem},
  author={Kuhn, Harold W},
  journal={Naval Research Logistics Quarterly},
  volume={2},
  number={1-2},
  pages={83--97},
  year={1955},
  publisher={Wiley Online Library},
  doi={10.1002/nav.3800020109}
}

\end{document}